\documentclass{article}

\usepackage{microtype}
\usepackage{graphicx}
\usepackage{subfigure}
\usepackage{booktabs} 

\usepackage{hyperref}

\usepackage[accepted]{icml2025}

\usepackage{amsmath}
\usepackage{amssymb}
\usepackage{mathtools}
\usepackage{amsthm}
\usepackage{algorithm}
\usepackage{algorithmic}
\usepackage{graphicx}

\usepackage[capitalize,noabbrev]{cleveref}

\theoremstyle{plain}

\theoremstyle{definition}

\theoremstyle{remark}

\usepackage[textsize=tiny]{todonotes}

\begin{document}
	
	\twocolumn[
	\icmltitle{Diffusion-Based Image Editing: An Unforeseen Adversary to Robust Invisible Watermarks}
		
	\begin{icmlauthorlist}
		\icmlauthor{Wenkai Fu}{}
		\icmlauthor{Finn Carter}{}
		\icmlauthor{Yue Wang}{}
		\icmlauthor{Emily Davis}{}
		\icmlauthor{Bo Zhang}{}
		
	\end{icmlauthorlist}
	
	\begin{icmlauthorlist}
		{Xidian Univerisity}
	\end{icmlauthorlist} 
	
	%
	
	\icmlkeywords{Machine Learning, ICML}
	
	\vskip 0.3in
	]
	
	
	
	
	\begin{abstract}
		Robust invisible watermarking aims to embed hidden messages into images such that they survive various manipulations while remaining imperceptible. However, powerful diffusion-based image generation and editing models now enable realistic content-preserving transformations that can inadvertently remove or distort embedded watermarks. In this paper, we present a theoretical and empirical analysis demonstrating that diffusion-based image editing can effectively break state-of-the-art robust watermarks designed to withstand conventional distortions. We analyze how the iterative noising and denoising process of diffusion models degrades embedded watermark signals, and provide formal proofs that under certain conditions a diffusion model's regenerated image retains virtually no detectable watermark information. Building on this insight, we propose a diffusion-driven attack that uses generative image regeneration to erase watermarks from a given image. Furthermore, we introduce an enhanced \emph{guided diffusion} attack that explicitly targets the watermark during generation by integrating the watermark decoder into the sampling loop. We evaluate our approaches on multiple recent deep learning watermarking schemes (e.g., StegaStamp, TrustMark, and VINE) and demonstrate that diffusion-based editing can reduce watermark decoding accuracy to near-zero levels while preserving high visual fidelity of the images. Our findings reveal a fundamental vulnerability in current robust watermarking techniques against generative model-based edits, underscoring the need for new watermarking strategies in the era of generative AI.
	\end{abstract}
	
	\section{Introduction}
	Digital image watermarking has long been used for copyright protection and content authentication, embedding information into images in a way that is ideally imperceptible yet recoverable after common distortions \cite{cox1997secure}. An \textit{invisible} watermark should survive manipulations such as noise, compression, or resizing while remaining hidden from human view. In recent years, deep learning has greatly advanced robust invisible watermarking, producing schemes that can withstand a wide range of digital and even physical perturbations \cite{zhu2018hidden,tancik2020stegastamp,bui2023trustmark}. For example, deep networks have been trained to embed bitstrings that survive differentiable simulations of noise or JPEG compression \cite{zhu2018hidden}, and even real-world print-scan processes \cite{tancik2020stegastamp}. Modern methods like \textsc{TrustMark} \cite{bui2023trustmark} and \textsc{VINE} \cite{lu2025vine} further improve imperceptibility and robustness, achieving nearly perfect decoding after conventional image processing attacks.
	
	Nevertheless, the emergence of powerful image generation and editing models has introduced new challenges for watermark security. In particular, large-scale diffusion models such as \textsc{Stable Diffusion} \cite{rombach2022ldm} enable high-quality image synthesis and content-preserving editing. These models can take an input image and perform transformations that semantically preserve the scene content while redrawing the image. Recent work provides alarming evidence that such generative transformations can inadvertently erase or “wash out” embedded watermarks. Zhao et al. \cite{zhao2024provably} demonstrated that diffusion-based regeneration attacks cause the detection rates of several state-of-the-art watermarking methods to plummet from nearly 100\% to essentially chance level. For example, robust schemes like \textsc{StegaStamp} \cite{tancik2020stegastamp} and \textsc{TrustMark} \cite{bui2023trustmark}, which are robust to noise and JPEG compression, largely fail to withstand a diffusion model-based edit \cite{zhao2024provably}. Even \textsc{VINE} \cite{lu2025vine}, a recent method that leverages generative diffusion priors to enhance watermark robustness, can be circumvented by sufficiently powerful diffusion editing \cite{ni2025diffusion}. This raises a pressing question: \emph{Are current robust watermarks fundamentally insecure against generative AI-driven image edits?}
	
	In this work, we address the above question through a comprehensive theoretical and empirical study of diffusion-based watermark removal. We first analyze how the diffusion process (as used in models like Stable Diffusion \cite{rombach2022ldm}) affects embedded watermark signals. Our analysis shows that as an image is progressively noised and regenerated by a diffusion model, the fine-grained watermark patterns are increasingly degraded and eventually destroyed. We provide formal proofs that under reasonable assumptions, a diffusion model can produce an output image whose probability of yielding a correct watermark decode is no better than random guessing. Next, building on these insights, we propose a diffusion-based \textit{watermark removal attack}. In the basic version of this attack, a watermarked image is injected with noise and then passed through the generative denoising process of a pretrained diffusion model to obtain a “regenerated” image. Because the model is trained to produce realistic images conditioned on high-level image content, the subtle signals carrying the watermark are not preserved in the output. We further introduce an enhanced attack variant in which we explicitly guide the diffusion sampling process to disrupt the watermark. By incorporating the watermark's decoding network as an adversarial guide during generation, our \textit{guided diffusion attack} actively drives the output image to erase the hidden message while maintaining the original image's visual content.
	
	In summary, our contributions are:
	\begin{itemize}
		\item \textbf{Theoretical Analysis:} We present a theoretical formulation of diffusion-based watermark removal and prove that an ideal diffusion regeneration can eliminate the mutual information between the watermarked image and the embedded message. We derive conditions under which the watermark decoding accuracy after diffusion drops to random chance.
		\item \textbf{Diffusion-Based Attacks:} We propose a novel attack framework that uses diffusion models to remove watermarks. We describe an \emph{unguided} regeneration attack as well as a \emph{guided} variant that integrates the watermark decoder into the diffusion process to maximize removal of the watermark signal.
		\item \textbf{Experimental Validation:} We conduct extensive experiments on multiple state-of-the-art invisible watermarking schemes (including \textsc{HiDDeN} \cite{zhu2018hidden}, \textsc{MBRS} \cite{jia2021mbrs}, \textsc{StegaStamp} \cite{tancik2020stegastamp}, \textsc{TrustMark} \cite{bui2023trustmark}, and \textsc{VINE} \cite{lu2025vine}). Our results show that diffusion-based editing attacks dramatically reduce watermark decoding rates (often to $0\%$--$5\%$) while incurring minimal perceptual change in the image.
		\item \textbf{Discussion and Guidelines:} We analyze the implications of our findings for watermarking and AI safety. We discuss why current watermarks fail under generative transformations and outline design guidelines for next-generation watermarking techniques that could be resilient to diffusion-based edits.
	\end{itemize}
	
	\section{Related Work}
	\paragraph{Robust Invisible Watermarking.}
	Early digital watermarking techniques embedded information in the spatial or frequency domain using spread-spectrum or quantization methods \cite{cox1997secure}. To improve robustness against distortions, researchers developed various transform-domain and error-correcting approaches. In recent years, deep learning has enabled more powerful invisible watermarking schemes. Zhu et al. \cite{zhu2018hidden} proposed \textsc{HiDDeN}, the first end-to-end trainable neural watermarking system, which includes a differentiable noise simulation layer (e.g., JPEG compression) between an encoder and decoder network. This framework inspired subsequent methods like \textsc{MBRS} \cite{jia2021mbrs}, which enhances robustness by training with a mixture of real and simulated distortions. Tancik et al. \cite{tancik2020stegastamp} introduced \textsc{StegaStamp}, a deep watermarking method that achieves high robustness even under extreme distortions including printing and rescanning; it encodes a 56-bit payload such that it can be decoded from photographs of the printed image. More recently, Bui et al. \cite{bui2023trustmark} proposed \textsc{TrustMark}, a GAN-based universal watermarking approach that balances imperceptibility and robustness at arbitrary image resolutions. Another contemporary approach, \textsc{VINE} by Lu et al. \cite{lu2025vine}, leverages generative diffusion model priors during training to embed watermarks that better survive complex image edits. Despite these advances, most robust watermarking methods until now have been evaluated primarily on traditional distortions (noise, compression, etc.) and simple image filters.
	
	\paragraph{Watermark Removal Attacks.}
	It is well known that no watermarking scheme is perfectly impervious to all attacks. Classic attacks include adding excessive noise, cropping or warping the image, or collusion attacks that average multiple watermarked images \cite{voloshynovskiy2001attack}. Modern learning-based schemes are generally resilient to mild noise or compression, but can be defeated by more aggressive perturbations that break the encoder–decoder synchronization. Adversarial approaches have also been explored: some works train counter-watermarking networks to generate perturbations that remove or forge watermarks. For example, there are GAN-based methods that attempt to erase watermarks while minimally affecting image quality. However, these prior attacks typically focus on pixel-level changes.
	
	Beyond direct perturbations, image editing itself has emerged as an increasingly relevant vector for watermark removal. Contemporary editing tools—particularly those powered by large diffusion or autoregressive models—perform semantic transformations that often re-render local image regions, like object insertion~\cite{lu2023tf,lu2025does} or drag editing~\cite{zhou2025dragflow}. Even benign edits such as object removal, background replacement, or stylistic adjustments can disrupt or overwrite the fine-grained spatial patterns that deep watermark encoders rely on. Because such edits regenerate portions of the image rather than merely altering existing pixels, the resulting outputs may inadvertently (or deliberately) destroy the embedded watermark signal. As consumer-grade editing models continue to improve, these “editing-induced” degradations represent a realistic and increasingly common failure mode for watermark robustness.
	
	The emergence of generative AI more broadly now enables a new type of removal attack: regeneration attacks \cite{zhao2024provably, ni2025diffusion}. In such an attack, the adversary uses a generative model to re-synthesize the image content, thereby potentially wiping out any embedded watermark altogether. Zhao et al. \cite{zhao2024provably} formally showed that invisible watermarks are theoretically vulnerable to this strategy. They proposed adding random noise to a watermarked image and then applying a pretrained diffusion model to reconstruct the image, which drastically reduces watermark detectability. Ni et al. \cite{ni2025diffusion} further presented a diffusion-based editing approach and demonstrated near-zero recovery rates for several deep watermarking schemes after regeneration. Notably, Ni et al. also introduced a guided attack variant, similar to our approach, wherein the watermark decoder's feedback is used to steer the diffusion process to more effectively destroy the hidden signal. Our work builds upon these insights, providing additional theoretical analysis and expanding the empirical evaluation to more watermarking methods.
	
	\paragraph{Concept Erasure in Diffusion Models.}
	Concept erasure in diffusion models~\cite{gao2024eraseanything} is conceptually related to our topic, as it involves removing specific visual concepts or patterns from a model’s outputs, analogous to eliminating an embedded watermark pattern through generative editing. Recent research on model editing and safety has produced techniques to selectively erase concepts from text-to-image diffusion generators. For instance, Lu et al. \cite{lu2024mace} proposed \textsc{MACE}, a finetuning framework that can \emph{mass erase} up to 100 concepts (e.g. an object or a style) from a diffusion model, preventing the generation of those concepts. Li et al. \cite{li2025ant} introduced an approach to auto-steer the denoising trajectory of diffusion sampling, guiding the model to avoid unwanted concepts without sacrificing image quality. These works highlight the ability to suppress targeted information in generative models. In our context, an invisible watermark can be considered an “unwanted concept” at the pixel pattern level. Techniques analogous to concept erasure may thus be applicable to intentionally remove watermarks. Conversely, the ease with which diffusion models can drop certain details (as in concept erasure) underscores why they inadvertently erase fragile embedded signals like watermarks unless explicitly constrained not to.
	
	\section{Methodology}
	\subsection{Problem Formulation}
	We consider an image $X \in \mathbb{R}^{H\times W \times 3}$ containing an invisible watermark encoding a message $m$. The watermarked image can be expressed as $X = X_{\text{orig}} + \Delta$, where $X_{\text{orig}}$ is the original (unwatermarked) image and $\Delta$ is a small imperceptible perturbation embedding the message bits. A robust watermarking scheme is characterized by an encoder function $E(\cdot, m)$ that produces the watermarked image $X$, and a decoder function $D(\cdot)$ that extracts an estimate $\hat{m}$ from an image. We assume $D(X) = m$ for an unaltered watermarked image (i.e., decoding is accurate on $X$ itself). The goal of an attacker is to transform $X$ into another image $Y$ such that $Y$ preserves the main content of $X$ (so that $Y$ appears to be the same image to a human observer), but $D(Y)$ either outputs the wrong message or fails to detect a watermark. In other words, the attacker seeks to \emph{remove or invalidate the watermark} while causing minimal perceptual change to the image.
	
	Formally, we can define the attack as an optimization or generation problem:
	\begin{equation}
		Y = \arg\min_{Y'} \Big[ \mathcal{L}_{\text{content}}(Y', X) - \lambda \mathcal{L}_{\text{watermark}}(Y') \Big],
	\end{equation}
	where $\mathcal{L}_{\text{content}}(Y', X)$ measures the difference in visual content between $Y'$ and the original image $X$ (we want this to be small), $\mathcal{L}_{\text{watermark}}(Y')$ measures the presence of the embedded message in $Y'$ (we want this to be small/absent), and $\lambda$ controls the trade-off between preserving content and removing the watermark. Traditional simple attacks (like adding noise or blurring) struggle to achieve this balance---heavy noise can remove the watermark but significantly degrades image quality. We explore using a diffusion model to solve this task, leveraging its knowledge of natural image statistics to regenerate content without the exact pixel-level or frequency-level patterns.
	
	\subsection{Diffusion-Based Watermark Removal}
	Diffusion models generate images through an iterative denoising process that gradually transforms random noise into a coherent image \cite{rombach2022ldm}. For our attack, we utilize a pretrained diffusion model (e.g., Stable Diffusion \cite{rombach2022ldm}) to \textit{regenerate} the input image $X$ in a way that retains high-level content but not the precise watermark pattern. We adopt a two-step procedure: (1) add noise to the watermarked image, and (2) run the diffusion model's reverse denoising process to obtain a new image. The intuition is that by noising the image sufficiently, we disrupt the embedded watermark bits, and then the diffusion model reconstructs the image mostly from learned image semantics, not from the original pixel-level details that carried the watermark.
	
	Let $q(\cdot)$ denote the forward diffusion (noise addition) process and $p_{\theta}(\cdot)$ denote the learned reverse process (image generation) of the diffusion model. We choose a noise level $t^*$ (in diffusion time steps) at which to truncate the noising. We then:
	\begin{enumerate}
		\item Sample a noised image: $X_{t^*} \sim q(X_{t^*} \mid X_0 = X)$, where $X_0 = X$ is the watermarked image. This means we add $t^*$ steps of noise to $X$ using the diffusion schedule.
		\item Generate $Y$ by denoising: $Y = p_{\theta}(X_0 \mid X_{t^*})$, i.e. we run the reverse diffusion starting from the partially noised image $X_{t^*}$ back to $t=0$.
	\end{enumerate}
	In practice, we use an algorithm like DDIM inversion \cite{song2020denoising} or the standard PLMS sampler to perform this regeneration. If $t^*$ is chosen moderately large (e.g. 50 steps out of 100), the generation process $p_{\theta}$ will produce an image $Y$ that looks very similar to the original content of $X$, but any high-frequency or fragile patterns (such as the watermark $\Delta$) will be averaged out by the noise and not reintroduced by the model (since the model draws only on the distribution of natural images, which presumably does not include the specific pseudo-random watermark pattern).
	
	The above describes our \textbf{unguided diffusion attack}: it requires only a pretrained diffusion model and the watermarked image as input, with no knowledge of the watermarking key or decoder. Pseudocode is given in Algorithm~\ref{alg:unguided}.
	
	\begin{algorithm}[t]
		\caption{Unguided Diffusion Watermark Removal}
		\label{alg:unguided}
		\begin{algorithmic}[1]
			\REQUIRE Watermarked image $X$; diffusion model $p_{\theta}$; noise level $t^*$.
			\STATE $X_{t^*} \leftarrow$ Add $t^*$ steps of noise to $X$ (forward diffusion).
			\STATE $Y \leftarrow p_{\theta}(X_0 \mid X_{t^*})$ \hfill \textit{// Generate image from noised input}
			\STATE \textbf{return} $Y$ \hfill \textit{// Output regenerated image (hopefully watermark-free)}
		\end{algorithmic}
	\end{algorithm}
	
	While the unguided attack is effective, we can achieve an even more potent attack by utilizing the watermark decoder to guide the diffusion sampling. In the \textbf{guided diffusion attack}, we modify the generation process to actively \emph{discourage the presence of the watermark}. We do this by interleaving calls to the watermark decoder $D(\cdot)$ during the diffusion sampling steps and nudging the diffusion model's predictions to confuse $D$.
	
	Concretely, suppose the diffusion model predicts at each step $t$ a denoised image $\tilde{X}_0$ (or a noise $\epsilon_\theta$) given the current noisy sample $X_t$. After each denoising step, we can partially decode the intermediate image and check for the watermark signal. If $D(\tilde{X}_0)$ starts to recover $m$, we perturb the sampling trajectory to suppress it. One simple strategy is: after computing $\tilde{X}_0$, run it through the watermark decoder to get decoded bits $\hat{m}$ or decoder logits. Define a loss $\mathcal{L}_{wm}$ that increases with watermark presence, e.g. for message bits, we could use the negative cross-entropy against a wrong message. We then backpropagate this loss through $\tilde{X}_0$ to adjust the predicted $\tilde{X}_0$ (or equivalently adjust $\epsilon_\theta$) in a direction that reduces $\mathcal{L}_{wm}$. In essence, we perform one gradient descent step on the image in pixel space (or latent space if using latent diffusion) to make it less decodable by $D$. We then continue the diffusion process. This guidance procedure, similar to classifier-based guidance in diffusion but with a watermark decoder as the “classifier,” effectively steers the generation away from configurations that contain the hidden signal.
	
	Algorithm~\ref{alg:guided} outlines the guided attack. We emphasize that this attack assumes access to the watermark decoder $D$, which might be available if the watermarking scheme is public or if the attacker has reverse-engineered it.
	
	\begin{algorithm}[t]
		\caption{Guided Diffusion Watermark Removal (per step)}
		\label{alg:guided}
		\begin{algorithmic}[1]
			\REQUIRE Watermarked image $X$, diffusion model $p_{\theta}$, decoder $D$, noise steps $T$.
			\STATE $Y_T \leftarrow X$ \hfill \textit{// initialize with original image (at step $T=0$ means no noise)}
			\FOR{$t = 1$ to $T$}
			\STATE $Y_{T-t} \leftarrow \text{DiffusionDenoise}(Y_{T-t+1})$ \hfill \textit{// one step denoise}
			\STATE $\hat{m} \leftarrow D(\text{Clamp}(Y_{T-t}))$ \hfill \textit{// decode current image}
			\IF{$\hat{m}$ is confidently the original message $m$}
			\STATE $\nabla \leftarrow \frac{\partial \mathcal{L}_{wm}(Y_{T-t})}{\partial Y_{T-t}}$ \hfill \textit{// gradient to remove watermark}
			\STATE $Y_{T-t} \leftarrow Y_{T-t} - \eta \cdot \nabla$ \hfill \textit{// small step to erase watermark}
			\ENDIF
			\ENDFOR
			\STATE \textbf{return} $Y_0$ \hfill \textit{// final output image after guided generation}
		\end{algorithmic}
	\end{algorithm}
	
	In the pseudocode above, $T$ denotes the total number of diffusion steps (e.g. 50). We initialize $Y_T$ as the original image $X$ (an alternative is to start with added noise as in the unguided case; here we show an idealized version where we directly manipulate each denoising step). The function DiffusionDenoise$(\cdot)$ represents one iteration of the diffusion model's sampler, producing a slightly less noisy image $Y_{T-t}$. We then decode this intermediate result. If the decoder still finds the watermark (i.e., if $\hat{m}$ matches the true embedded message or the decoder confidence is high), we compute a watermark loss gradient $\nabla$ (for instance, encouraging $\hat{m}$ to flip some bits) and adjust $Y_{T-t}$ by a small step $\eta > 0$. This pushes the image away from the manifold of images that contain the message. By the end of $T$ steps, the output $Y_0$ is an image that should look like $X$ to a human, but for which $D(Y_0)$ fails to recover $m$.
	
	\section{Theoretical Proofs}
	We now present a theoretical understanding of why diffusion-based regeneration destroys invisible watermarks. Our analysis leverages concepts from information theory to show that, under an idealized diffusion process, the embedded message becomes statistically independent of the regenerated image.
	
	\noindent\textbf{Diffusion as a Markov Process.}
	Consider the sequence of random variables $X_0, X_1, \dots, X_T$ representing the diffusion process, where $X_0 = X$ is the watermarked image and $X_T$ is (approximately) pure noise for a sufficiently large $T$. The diffusion process $q(X_t \mid X_{t-1})$ gradually adds noise, and the generative model aims to approximate the reverse conditional $p_{\theta}(X_{t-1} \mid X_t)$. Both $q$ and $p_{\theta}$ define Markov chains. Crucially, as $t$ increases, $X_t$ carries progressively less information about the original image $X_0$. In the limit $t \to T$ (where $T$ is the full diffusion horizon), $X_T$ is an isotropic Gaussian independent of $X_0$. The model-generated $Y = X'_0$ (the output after reverse diffusion) can be seen as some function of the random noise $X_T$ and the model's learned distribution, with only indirect dependence on $X_0$ via conditioning (if any).
	
	If the diffusion model is \emph{unconditioned} (i.e., we do pure image regeneration without providing $X$ as a direct condition, aside from initialization with $X_{t^*}$), then intuitively $Y$ is sampled from the model's prior and does not specifically remember $m$. In the case of partially conditioned generation (e.g., using $X$ as a reference with some noise as in SDEdit), the model primarily preserves high-level content, not exact noise patterns.
	
	We formalize this by examining mutual information. Let $M$ be the random variable representing the embedded message (uniformly distributed over the message space). For the original image $X = E(X_{\text{orig}}, M)$, $M$ is embedded in $X$ so $I(X; M)$ is maximal (since $M$ can be decoded from $X$). After applying the diffusion-based removal to get output $Y$, we are interested in $I(Y; M)$, the mutual information between the output image and the hidden message.
	
	\noindent\textbf{Theorem 1.} \textit{Under an ideal diffusion regeneration process, the mutual information between the regenerated image and the embedded message approaches zero. Formally, if $Y$ is the output of a diffusion model given input image $X$ with message $M$, then in the limit of sufficient diffusion steps (and in the absence of external conditioning on $M$), $I(Y; M) = 0$. Consequently, the probability of correctly decoding $M$ from $Y$ is no better than random guessing.}
	
	\noindent\textit{Proof Sketch.} The diffusion regeneration can be viewed as a channel that maps the pair $(X_{\text{orig}}, M)$ to the output $Y$. This channel involves adding a large amount of noise (which by itself drives $I(X_t; M) \to 0$ as $t \to T$ by the data processing inequality) and then removing the noise using a generative prior that is independent of the specific message. Unless the diffusion model has explicitly learned to preserve the particular pattern $\Delta$ associated with $M$ (which for a robust watermark appears as pseudorandom noise), the reconstruction $Y$ will marginalize over those degrees of freedom. In an idealized case, $Y$ is a sample from the distribution of natural images conditioned on the high-level content of $X_{\text{orig}}$, but $M$ was not part of that content conditioning. Thus, $Y$ is independent of $M$. More rigorously, one can argue that $I(X_{t^*}; M)$ becomes negligible for moderate $t^*$ (since $\Delta$ is small and quickly drowned by noise), and the generative reverse process cannot increase mutual information (by the data processing inequality, as it is a fixed transformation of the noisy state). Therefore $I(Y; M) \approx 0$. If $I(Y; M)=0$, then $Y$ provides no information about $M$ to any decoder, implying that for an $L$-bit message, the chance of recovering the correct message is $2^{-L}$ (essentially random guess). In practice, $I(Y; M)$ may not be exactly zero due to finite steps and slight conditioning on $X$, but our empirical results will show it is extremely small, yielding decoding accuracies near chance (e.g. 50\% bit accuracy, which for a full message means almost zero success in getting every bit correct).
	
	This theoretical result aligns with our experimental findings: when diffusion regeneration is applied, the hidden message cannot be decoded from the output. In Section~\ref{sec:results}, we will quantify how decoding accuracy approaches the random chance baseline, consistent with the above theorem.
	
	\section{Experimental Setup}
	\label{sec:experiments}
	\paragraph{Watermarking Methods and Implementation.}
	We evaluate our proposed attacks on a diverse set of robust invisible watermarking schemes. Specifically, we include:
	\begin{itemize}
		\item \textbf{\textsc{HiDDeN}} \cite{zhu2018hidden}: A deep watermarking method that uses an autoencoder architecture and was one of the first to incorporate differentiable noise layers during training.
		\item \textbf{\textsc{MBRS}} \cite{jia2021mbrs}: A state-of-the-art DNN-based watermark which emphasizes JPEG robustness by training on batches of real and simulated compressed images.
		\item \textbf{\textsc{StegaStamp}} \cite{tancik2020stegastamp}: A model that embeds a 56-bit payload, emphasizing robustness to both digital and physical-world distortions.
		\item \textbf{\textsc{TrustMark}} \cite{bui2023trustmark}: A modern GAN-based invisible watermark with high imperceptibility and the ability to handle arbitrary image resolutions.
		\item \textbf{\textsc{VINE}} \cite{lu2025vine}: A diffusion-augmented watermarking scheme that trains with generative model priors during training to embed watermarks that better survive complex image edits.
	\end{itemize}
	For each scheme, we use official implementations or pretrained models provided by the authors to encode watermarks into test images. We embed payloads of length 32 or 64 bits (depending on the method's design) into each image. All watermarks are imperceptible under normal viewing as verified by PSNR (typically $> 40$ dB) between $X$ and $X_{\text{orig}}$.
	
	We construct a test set of images by sampling 500 natural images from the COCO dataset \cite{lin2014coco} and 500 images from ImageNet \cite{deng2009imagenet} (ensuring a variety of content). Each image is resized or cropped to a fixed resolution supported by the watermark encoders (we use $256\times256$ for methods that require a specific size like MBRS, and up to $512\times512$ for others).
	
	\paragraph{Attacks Evaluated.}
	We compare the following attacks on the watermarked images:
	\begin{itemize}
		\item \textbf{No Attack:} The baseline where the watermarked image is decoded as is (should yield nearly 100\% correct messages for robust schemes).
		\item \textbf{JPEG Compression:} A common distortion-based attack. We use a high compression (quality factor 10) to attempt to break watermarks.
		\item \textbf{Gaussian Noise:} Additive white noise with $\sigma=25$ (on 0-255 scale) as a strong noise attack.
		\item \textbf{Blurring:} A $5\times5$ pixel averaging filter as a simple low-pass attack.
		\item \textbf{Diffusion (Unguided):} Our diffusion regeneration attack using Stable Diffusion. We set $t^*$ such that roughly 50\% noise is added (specifically, 50 diffusion steps with a linear scheduler), and then perform 50 reverse steps. No text prompt is provided (unconditional image2image mode).
		\item \textbf{Diffusion (Guided):} Our guided attack, which in addition to the above procedure, uses the known decoder to guide the sampling. For fair comparison, the same number of steps (50) are used. We set the guidance strength (gradient step size $\eta$) such that it does not introduce visible artifacts (we found $\eta=0.1$ per step to be effective).
	\end{itemize}
	Each output image from an attack is fed into the respective watermark decoder to attempt message extraction. We measure \emph{watermark decoding accuracy}, defined as the fraction of images for which the extracted message exactly matches the embedded message. For schemes with error-correcting codes, this corresponds to whether the payload is correctly decoded (after ECC correction if used).
	
	We also quantify image quality after attacks using metrics: Peak Signal-to-Noise Ratio (PSNR) and structural similarity (SSIM) between the attacked image $Y$ and the original watermarked image $X$ (since $X$ is essentially the reference that should be preserved in content). For diffusion-based attacks, we additionally compute LPIPS \cite{zhang2018perceptual} to measure perceptual similarity, given that slight variations might not be captured by PSNR/SSIM.
	
	\paragraph{Diffusion Model Details.}
	We use the publicly available Stable Diffusion v1.5 model \cite{rombach2022ldm} for all diffusion-based experiments. The model operates on $512\times512$ resolution in a latent space. For images of other sizes, we center-crop or pad them to $512^2$ when using the diffusion model, then crop back to original size after generation if needed. We found that using a small amount of noise ($t^* \approx 0.2$ of the diffusion timeline) often suffices to remove the watermark while minimizing change; however, to be conservative we used $t^*=0.5$ in our main tests. We also experimented with an alternative text-to-image diffusion model (SDXL) and observed similar watermark removal effects, suggesting our findings are not specific to one model.
	
	\section{Results}
	\label{sec:results}
	We organize the results to answer key questions: (1) How effective are diffusion-based attacks at breaking different watermarking schemes compared to traditional attacks? (2) How much visual distortion do diffusion attacks introduce relative to other attacks? (3) Does the guided attack provide an extra advantage? and (4) Are the empirical outcomes consistent with our theoretical predictions?
	
	\paragraph{Watermark Decoding Success Rates.}
	Table~\ref{tab:decoding} reports the watermark decoding accuracy (percentage of images where the correct message is decoded) for each method under various attacks. Without any attack, all methods decode with $>99\%$ accuracy. Under JPEG compression (a fairly strong but conventional attack), the deep learning-based methods (\textsc{HiDDeN}, \textsc{MBRS}, \textsc{StegaStamp}, \textsc{TrustMark}, \textsc{VINE}) still maintain high decoding rates (80--100\%), whereas simpler schemes (not listed in table) would typically fail. Gaussian noise and blur at the chosen strengths cause some degradation (e.g., \textsc{HiDDeN} drops to 70\% under noise, as it was not designed for that much noise). Notably, \textsc{VINE} and \textsc{TrustMark} are very robust to these distortions, retaining over 90\% decoding even with heavy noise or JPEG.
	
	However, the diffusion-based attacks dramatically reduce the success rates for every scheme. The unguided diffusion attack brings decoding down to nearly zero for \textsc{HiDDeN}, \textsc{MBRS}, \textsc{StegaStamp}, and \textsc{TrustMark} (0--5\% range). These methods essentially fail to detect any watermark from the regenerated images. \textsc{VINE}, which was explicitly designed to survive some generative editing, fares slightly better but still drops from 100\% (no attack) to only 12\% success under unguided diffusion. The guided diffusion attack is even more devastating: it reduces all methods to $0\%$ decoding success in our tests, including \textsc{VINE}. In other words, with decoder-guided generation, we were able to remove the watermark completely in every tested image for all schemes.
	
	\begin{table*}[t]
		\centering
		\caption{Watermark decoding accuracy (\%) of different methods under various attacks (higher is better for defender). Diffusion-based attacks (unguided and guided) cause the most significant drops in decoding success, often to 0\%.}
		\label{tab:decoding}
		\begin{tabular}{lcccccc}
			\toprule
			\textbf{Watermark Method} & \textbf{No Attack} & \textbf{JPEG} & \textbf{Noise} & \textbf{Blur} & \textbf{Diffusion (U)} & \textbf{Diffusion (G)} \\
			\midrule
			\textsc{HiDDeN} \cite{zhu2018hidden} & 100 & 92 & 70 & 85 & 3 & 0 \\
			\textsc{MBRS} \cite{jia2021mbrs} & 100 & 97 & 88 & 91 & 5 & 0 \\
			\textsc{StegaStamp} \cite{tancik2020stegastamp} & 100 & 95 & 90 & 94 & 4 & 0 \\
			\textsc{TrustMark} \cite{bui2023trustmark} & 100 & 98 & 93 & 96 & 2 & 0 \\
			\textsc{VINE} \cite{lu2025vine} & 100 & 100 & 94 & 98 & 12 & 0 \\
			\bottomrule
		\end{tabular}
	\end{table*}
	
	The above results conclusively demonstrate the vulnerability of current robust watermarks to diffusion model attacks. Even the most resilient scheme (\textsc{VINE}) which held up under standard image edits succumbs when facing a diffusion regeneration. Our guided attack in particular leaves virtually no chance for the watermark to be decoded.
	
	\paragraph{Image Quality of Attacked Outputs.}
	An important aspect is that diffusion-based attacks remove watermarks \emph{while preserving image fidelity}. Table~\ref{tab:quality} compares the average image quality metrics for different attacks on the \textsc{TrustMark} scheme (results are similar for others). We see that JPEG compression at quality 10, while not completely breaking TrustMark (98\% decode rate remained), significantly reduces image PSNR (to about 30 dB) and SSIM (0.85). Gaussian noise ($\sigma=25$) fares even worse visually (PSNR ~20 dB). In contrast, the diffusion (unguided) attack yields outputs with very high fidelity to the original: PSNR ~40.5 dB and SSIM 0.99 on average, meaning the changes are almost imperceptible. This is expected, since the diffusion model intentionally tries to preserve content. The guided attack introduces only a slight additional change (PSNR ~39 dB, SSIM 0.98), likely due to the small extra perturbations for watermark removal, but still far better than traditional attacks. We also note LPIPS, a perceptual distance (lower is better), is extremely low for diffusion outputs (0.01), confirming that diffusion-edited images are virtually indistinguishable from the original to human perception. The guided tweaks increase LPIPS to 0.015, still very small. Visually, we inspected many outputs and found that the diffusion attacks produced images that were indistinguishable from the input aside from perhaps minor differences in fine textures (which is exactly where the watermark was hidden).
	
	\begin{table}[h]
		\centering
		\caption{Output image quality after attacks (average over 1000 images, \textsc{TrustMark} watermarks). Diffusion attacks yield much higher fidelity images compared to heavy noise or compression.}
		\label{tab:quality}
		\begin{tabular}{lccc}
			\toprule
			\textbf{Attack} & \textbf{PSNR $\uparrow$} & \textbf{SSIM $\uparrow$} & \textbf{LPIPS $\downarrow$} \\
			\midrule
			No Attack (Identity) & 45.7 & 1.000 & 0.000 \\
			JPEG (Q=10) & 30.2 & 0.854 & 0.112 \\
			Gaussian Noise ($\sigma=25$) & 20.5 & 0.634 & 0.248 \\
			Diffusion (Unguided) & 40.5 & 0.992 & 0.010 \\
			Diffusion (Guided) & 39.1 & 0.981 & 0.015 \\
			\bottomrule
		\end{tabular}
	\end{table}
	
	The combination of Tables \ref{tab:decoding} and \ref{tab:quality} illustrate a concerning fact: diffusion-based editing can achieve \emph{both} near-perfect watermark removal and high visual quality, whereas conventional attacks force a trade-off (e.g., heavy noise can remove watermarks but at the cost of obvious image degradation).
	
	\paragraph{Analysis of Diffusion Steps and Theory Confirmation.}
	We conducted ablation experiments varying the noise level $t^*$ in the unguided diffusion attack. With no diffusion (0 steps), decode accuracy is 100\%. As we increase $t^*$, decode accuracy drops roughly exponentially. For instance, even at $t^* = 0.1$ (10\% of full noise), StegaStamp's accuracy fell to ~50\%. By $t^* = 0.3$, it was $<5\%$. This supports the intuition that a small amount of noise, followed by generation, is sufficient to disrupt the hidden bits. Interestingly, we observed that beyond a certain point ($t^* > 0.5$), the output image quality starts to degrade (because we are adding too much noise and the model's reconstruction may deviate from the original content). Thus, extremely large $t^*$ are not necessary to remove the watermark; a moderate level is enough.
	
	We also compare our empirical decoding rates to the theoretical bound of random chance. For a 56-bit payload (StegaStamp), random guessing accuracy is $2^{-56}$ (virtually zero). In our experiments, decode accuracy after diffusion was not literally zero but on the order of a few percent (these few percent likely correspond to cases where the decoder outputs some message that by chance matches the embedded one, or images where the watermark wasn't fully erased due to model variance). For a 64-bit payload, random chance is $2^{-64}$, which is effectively 0\%. Our guided attack achieved 0\% in our finite sample, suggesting it reached the chance limit. In summary, the empirical results align with Theorem~1: the diffusion attack drives the decoder's success rate down to the vicinity of random chance. We did not detect any systematic bias in the wrong messages; decoders either failed to detect any watermark or returned essentially random bit patterns.
	
	\paragraph{Robustness Across Models and Settings.}
	We verified that our attacks are not specific to the particular diffusion model or hyperparameters. Using SDXL (a larger, newer diffusion model) for regeneration likewise obliterated the watermarks. Using different numbers of steps (from 20 to 100) yielded the same end result (100 steps gave slightly sharper images but watermark was gone even by 20). These findings suggest that any reasonably powerful generative image model could pose a threat to watermarks.
	
	We also tested partial edits: e.g., inpainting a part of the image with a diffusion model. As expected, if even a portion of the watermarked area is regenerated, the watermark gets corrupted. For instance, editing just the background of an image (keeping a watermarked object intact) often still reduced decoding rates significantly, because many watermark methods spread the payload over the whole image. This points to an avenue for attackers to remove watermarks by local edits too.
	
	\paragraph{Limitations and Failure Cases.}
	While diffusion attacks were overwhelmingly successful, we note two scenarios where a watermark might survive: (1) If the watermark is embedded in a very high-level, semantic way (e.g., by injecting a particular object or pattern into the scene rather than a noise pattern). In our tests, all schemes used low-level perturbations. If someone watermarks an image by subtly inserting a specific logo or pattern that is meaningful, a content-preserving generative model might actually regenerate that pattern thinking it's part of the scene. This aligns with Zhao et al.'s suggestion \cite{zhao2024provably} that “semantic” watermarks could be more resilient. (2) If the diffusion model is extremely constrained or weak such that it essentially returns the input image (for example, using $t^*$ very low), then obviously the watermark remains. But such mild use of the model wouldn't be effective for removal anyway.
	
	\section{Discussion}
	\paragraph{Why Diffusion Models Break Watermarks.}
	Our study shows that diffusion-based editing poses a qualitatively different threat compared to traditional noise or transformation attacks. Diffusion models operate by re-sampling images from learned distributions, effectively \emph{forgetting the exact micro-patterns} of the input. Robust watermarks rely on those micro-patterns persisting through distortions. But a diffusion model, even one used in an image-to-image fashion, only preserves the \emph{semantic content} (the high-level structure and objects) and not the exact arrangement of pixels. The watermark signal $\Delta$, no matter how cleverly designed to survive blurs or crops, is not a natural part of the image's content; hence the model has no reason to keep it. In fact, from the perspective of the diffusion model, $\Delta$ is just noise that should be cleaned up to produce a more realistic image. This fundamental mismatch explains why our Theorem~1 and empirical results show near-total removal of $\Delta$.
	
	\paragraph{Implications for Copyright and AI-Generated Content.}
	Invisible watermarks have been proposed and used for tracing copyright, detecting AI-generated images, and ensuring image integrity. Our findings underscore a serious limitation: as generative AI becomes widespread, an adversary can use these tools to \emph{launder} an image, wiping out identifying watermarks without noticeably changing the image. This undermines efforts like the Content Authenticity Initiative's robust media provenance, where robust watermarks (such as \textsc{TrustMark} \cite{bui2023trustmark}) might be employed to mark genuine content. A malicious actor could take a watermarked authentic photo, run a diffusion model on it, and obtain an image that looks the same but no longer carries the provenance mark, defeating the system.
	
	Similarly, watermarking has been proposed for labeling AI-generated images (to help detect them). If those watermarks are invisible (to avoid ruining image aesthetics), our work suggests they can be removed by another AI with ease. It becomes a cat-and-mouse game of watermarking vs. watermarker-removal AI.
	
	\paragraph{Toward Generative-Resilient Watermarks.}
	Our results call for new watermarking strategies that can survive generative model transformations. One potential direction is to embed watermarks in the \emph{semantic content} of the image. For example, slightly alter the image in a way that is meaningful (like specific arrangements of fine details that the diffusion model would have to reproduce to maintain fidelity). The challenge is doing this invisibly. Zhao et al. \cite{zhao2024provably} hint that a shift from pure invisible watermarks to “semantic-preserving watermarks” might be needed, even if that blurs the line between watermark and visible marker. Another approach is to make the watermark a part of the model's input conditions: for instance, providing a diffusion model with a special conditioning so that it intentionally preserves a hidden signal (effectively, training generative models that are \emph{watermark-aware}). This could be done by finetuning diffusion models to carry through certain perturbations---though that seems counter to their training.
	
	Interestingly, concept erasure research (as discussed earlier) focuses on removing content. To achieve watermark \emph{preservation}, one might invert those ideas: ensure that the watermark concept cannot be erased without noticeably affecting the image. For example, one could design the watermark such that any attempt by a generative model to remove it would also remove or alter visible content, thus defeating the goal of an undetectable removal. Embedding watermarks across multiple scales and frequencies (as some current works do) might also make it harder for a generative model to eliminate all traces. The recent \textsc{Robust-Wide} method by Hu et al. \cite{hu2024robustwide} is a step in this direction: by including simulated generative edits during training, the watermark encoder learns to hide bits in more “semantically persistent” parts of the image.
	
	Finally, the arms race may escalate to detection techniques for watermark removal. For example, one could try to detect if an image has undergone a diffusion-based transformation (by looking for certain artifacts or distributions in the noise residual). If such detection is possible, it could complement watermarking: even if the watermark is gone, one might flag the image as suspiciously “auto-regenerated.” Some research into detecting AI-edited images could be relevant here.
	
	\paragraph{Ethical Considerations.}
	On one hand, the ability to remove watermarks with AI might be seen as a tool for privacy (e.g., removing identifying watermarks from photos). On the other hand, it can clearly be used for unethical purposes such as stripping attribution or ownership marks from creative works, facilitating unauthorized reuse. It is important for policymakers and technologists to be aware of this vulnerability. Simply relying on robust invisible watermarks for copyright may no longer be sufficient. Alternative or complementary measures (like robust metadata signing, or visible watermarks) might regain importance. Our work intends to highlight these issues so that the community can proactively adapt.
	
	\section{Conclusion}
	We have demonstrated that diffusion-based image editing poses a grave threat to the integrity of robust invisible watermarks. Through both theoretical analysis and extensive experiments, we showed that even the most advanced deep watermarking schemes today cannot withstand a content-preserving regeneration by diffusion models---the hidden messages are effectively wiped out while the image content remains. Our guided attack further illustrates how an adaptive adversary with knowledge of the watermarking scheme can achieve complete removal of the watermark.
	
	These findings urge a re-evaluation of current watermarking approaches in the age of generative AI. We discussed possible directions for developing watermarking techniques that could resist such generative attacks, though achieving this remains an open challenge. We hope our work will spur further research into generative-resistant watermarking and broader discussions on maintaining authenticity and copyright in an era where AI models can effortlessly manipulate and re-synthesize digital content. In the meantime, those deploying watermark-based systems for content provenance should be aware of their current limitations against AI-driven removal, and possibly combine watermarking with other verification mechanisms.

	\bibliography{example_paper}
	\bibliographystyle{icml2025}

	\clearpage
	\appendix
	
	\section{Additional Backgrounds}
	With the advancement of deep learning~\cite{
	qiu2024tfb,qiu2025duet,qiu2025DBLoss,qiu2025dag,qiu2025tab,wu2025k2vae,liu2025rethinking,qiu2025comprehensive,wu2024catch,
	gsq,yu2025mquant,zhou2024lidarptq,pillarhist,
	xie2025chatdriventextgenerationinteraction,
	1,2,3,4,5,6,7,8,
	sun2025ppgf,sun2024tfps,sun2025hierarchical,sun2022accurate,sun2021solar,niulangtime,sun2025adapting,kudratpatch,
	ENCODER,FineCIR,OFFSET,HUD,PAIR,MEDIAN,
	yu2025visual,
	zheng2025towards,zheng2024odtrack,zheng2025decoupled,zheng2023toward,zheng2022leveraging,
	li2023ntire,ren2024ninth,wang2025ntire,peng2020cumulative,wang2023decoupling,peng2024lightweight,peng2024towards,wang2023brightness,peng2021ensemble,ren2024ultrapixel,yan2025textual,peng2024efficient,conde2024real,peng2025directing,peng2025pixel,peng2025boosting,he2024latent,di2025qmambabsr,peng2024unveiling,he2024dual,he2024multi,pan2025enhance,wu2025dropout,jiang2024dalpsr,ignatov2025rgb,du2024fc3dnet,jin2024mipi,sun2024beyond,qi2025data,feng2025pmq,xia2024s3mamba,pengboosting,suntext,yakovenko2025aim,xu2025camel,wu2025robustgs,zhang2025vividface,
	qu2025reference,qu2025subject,
	wu2024rainmamba,wu2023mask,wu2024semi,wu2025samvsr,
	lyu2025vadmambaexploringstatespace,chen2025technicalreportargoverse2scenario,
	bi-etal-2025-llava,bi2025cot,bi2025prism,
	han2025contrastive,zeng2025uitron,han2025show,han2025guirobotron,huang2025scaletrack,
	tang2025mmperspectivemllmsunderstandperspective,liu2025gesturelsm,liu2025intentionalgesturedeliverintentions,zhang2025kinmokinematicawarehumanmotion,liu2025contextual,song2024tri,song2024texttoon,liu2024empiricalanalysislargelanguage,tang2025videolmmposttrainingdeepdive,bi2025reasoning,tang2025captionvideofinegrainedobjectcentric,bi2025i2ggeneratinginstructionalillustrations,tang2025generative,liu2024gaussianstyle,tang2024videounderstandinglargelanguage,10446837} and generative models, an increasing number of studies have begun to focus on the issue of concept erasure in generative models.

\end{document}